\documentclass[aps,pra,reprint,superscriptaddress,twocolum]{revtex4-1}

\usepackage{amsmath}
\usepackage{amssymb}
\usepackage{hyperref}
\usepackage{graphicx}
\usepackage{epstopdf}
\usepackage{dcolumn}
\usepackage{bm}
\usepackage{xspace}
\usepackage{verbatim}
\usepackage{color}
\newcommand{\bra}[1]{\left\langle #1 \right|}
\newcommand{\ket}[1]{\left| #1 \right\rangle}

\hypersetup{colorlinks=true,linkcolor=blue,citecolor=blue, filecolor=blue,urlcolor=blue,breaklinks=true}

\begin{document}
\title{Enabling entanglement distillation via optomechanics}
\author{Victor Montenegro}
\email{vmontenegro@uestc.edu.cn}
\affiliation{Institute of Fundamental and Frontier Sciences,
University of Electronic Science and Technology of China, Chengdu, PR China}
\affiliation{Department of Physics and Astronomy, University College London, Gower Street, London WC1E 6BT, United Kingdom}

\author{Alessandro Ferraro}
\affiliation{Centre for Theoretical Atomic, Molecular, and Optical Physics, School of Mathematics and Physics, Queen's University, Belfast BT7 1NN, United Kingdom}

\author{Sougato Bose}
\affiliation{Department of Physics and Astronomy, University College London, Gower Street, London WC1E 6BT, United Kingdom}

\date{\today}

\begin{abstract}
Quantum networking based on optical Gaussian states, although promising in terms of scalability, is hindered by the fact that their entanglement cannot be distilled via Gaussian operations.  We show that optomechanics, integrable (on-chip) availability, and particularly the scope to measure the mechanical degree of freedom, can address this problem. Here, one of the optical modes  of a two-mode squeezed vacuum is injected into a single sided Fabry-P\'{e}rot cavity and non-linearly coupled to a mechanical oscillator. Afterwards, the position of the oscillator is measured using pulsed optomechanics and homodyne detection. We show that this measurement can supply non-Gaussian entangled states frequently enough to enable scalable entanglement distillation. Moreover, it can conditionally increase the initial entanglement under an optimal radiation-pressure interaction strength, which corresponds to an effective unsharp measurement of the photon number inside the cavity. We show how the resulting entanglement enhancement can be verified by using a standard teleportation procedure. 
\end{abstract}

\maketitle

\section{Introduction}

Recent experiments with quantum optics have demonstrated the generation of entanglement across up to one million of modes \cite{Furusawa2, Pfister, Fabre, Furusawa3},  thus offering unprecedented opportunities for quantum networking \cite{Kimble-Nature}. However, the states generated in these systems (Gaussian states of light) suffer from the drawback that entanglement distillation --- a pivotal primitive for long distance quantum communication  \cite{distillation} --- is not readily available. This is due to the fact that the interactions naturally occurring in these systems are Gaussian and  a ``no-go theorem'' prevents Gaussian operations to distill Gaussian entanglement \cite{no-go}. Some non-Gaussian element, which acts as a resource \cite{RTNG}, needs to be used to overcome this roadblock. In particular, optical methods involving non-Gaussian operations have been suggested \cite{Cirac, Fiurasek, browne, Walmsley, Datta, eisert1, eisert2}, with the dominant scheme relying on photon subtraction \cite{Dakna, Opatrny, Olivares, oli, ourj, mskim, dell, fiur, Navarrete, Kumar}. The implementation of such schemes is currently topical but remains challenging \cite{Wenger, Franzen, parigi, dong, Grangier, Furusawa, kurochkin, treps} with the rate of production of non-Gaussian states about a KHz. We introduce here an alternative based on hybrid opto-mechanical systems exploiting their natural non-Gaussian radiation-pressure interactions and the availability of a mechanical ``meter'' which can be measured efficiently.

\begin{figure}[t]
 \centering \includegraphics[scale = 0.8]{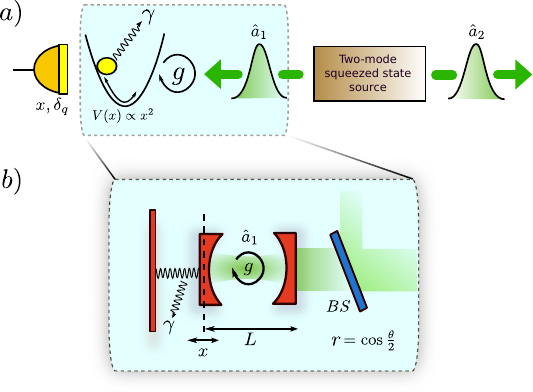}\caption{Concentration scheme for two-mode squeezed vacuum (TMSV) states. In a) we show the general idea, in which one mode $\hat{a}_1$ interacts with a damped ($\gamma$) mechanical harmonic oscillator with strength $g$. Subsequently, we proceed to measure the position ($x$) of the mechanical oscillator, thus increasing the initial TMSV entanglement. In general, we model the position measurement considering an ideal detector preceded by a beam splitter (BS) of transmissivity $\nu$, being $\delta_q = (1 - \nu)/(4\nu)$. In b) we have substituted the above scheme with a typical optomechanical setup, modeling the injection of the mode $\hat{a}_1$ into the cavity via BS of reflectivity $r$, and where the position of the mirror is to be measured by pulsed optomechanics \cite{vanner}.}\label{ab}
\end{figure}

Quantum optomechanics is opening up new avenues for the manipulation of optical states \cite{aspelmeyer1, aspelmeyer2}. The usage of optomechanical systems for teleportation and establishing Gaussian entangled states of distant systems have been studied (see, \textit{e.g.} Refs. \cite{Tombesi, Lloyd, zoller1, new-Tombesi}). However the key quantum communication enabling protocol of entanglement distillation has thus far been untouched in optomechanics as the majority of the applications considered a linearized (therefore Gaussian) interaction \cite{Maimaiti2015, Zheli2017}. On the other hand, the bare optomechanical radiation pressure interaction is non-Gaussian (trilinear) \cite{Law}. It typically entangles the mechanical and optical degrees of freedom \cite{Bose1999, Bouwmeester,Li} so that the mechanics can be measured to manipulate the state of light. As a matter of fact, this trilinear coupling is becoming physically significant in certain setups \cite{Painter, chan, murch, xuereb, Barclay} and drawing considerable attention \cite{Nunnenkamp1,Rabl, He, Liao1, Xu, Kronwald, Wei, Akram, Liao2}. A natural question thus arises: can the radiation pressure interaction enable entanglement distillation? Here we answer in the affirmative, showing that non-Gaussian entangled states of light can be produced at a sufficient rate to enable scalable entanglement distillation \cite{Datta} (see Section \ref{sec-enabling} for details). In particular, the entanglement of two mode squeezed vacua can be enhanced conditionally on ``snap-shot'' position detections of a mechanical oscillator --- via a methodology recently developed \cite{vanner, vanner2, aspelmeyer, Raizen,Novotny,Hendrik,Genoni1}. This snap-shot position measurement effectively measures the photon number in the cavity, as the optomechanical nonlinearity couples the photon number operator to the mechanical position \cite{fock-meter}. However, due to the weakness of the nonlinear optomechanical interaction, the photon number is effectively measured in a coarse grained manner through the snap-shot position measurement. Our proposal thus also illustrates that weak (in the sense of ``coarse-grained''/unsharp) measurements can be useful for enabling a quantum protocol.

The rest of the paper is structured as follows. In Section \ref{sec:sys-dyn}, we present the quantum optomechanical system in the nonlinear regime, where we analytically solved the standard master equation at zero temperature considering detrimental effects due to both optical injection, as well as the decoherence of the mechanical oscillator. In Section \ref{sec:ent-conc}, we show how to concentrate the initial optical modes via position measurement of the mechanical object. Moreover, we exhibit that by choosing an appropriate set of values, this entanglement concentration can be optimal. Next, in Section \ref{sec-tel}, to illustrate how useful is our generated state for quantum communication protocols, we show that when using our concentrated state, the quantum teleportation fidelity of an arbitrary coherent state is enhanced overall. In Section \ref{sec-enabling}, we justify how scalable entanglement distillation might also be constructed, a statement based on some schemes already relying on repeated production of non-Gaussian entangled states and linear optics. Section \ref{sec-exp}, aims to cover the experimental feasibility of our proposal. Finally, in Section \ref{sec-conclusions}, we outline the final remarks of our work.

\section{System Dynamics}\label{sec:sys-dyn}

Let us commence by considering two light-modes (with corresponding annihilation operators $\hat{a}_1$ and $\hat{a}_2$ satisfying $[\hat{a}_j, \hat{a}^\dag_j]=1$ for $j=1,2$) in a two-mode squeezed vacuum (TMSV)

\begin{equation}
\ket{\psi(0)}_{\mathrm{TMSV}} = \sqrt{1 - \lambda^2}\sum_{n=0}^\infty\lambda^n\ket{n,n}_{12}, \label{tmsv}
\end{equation}

with $\lambda = \mathrm{tanh} (s)$ and $s$ being the squeezing parameter. One light beam $(\hat{a}_1)$ is coupled to a mechanical harmonic oscillator, whereas mode $\hat{a}_2$ propagates freely [a general scheme is illustrated in Fig. (\ref{ab}-a)]. As said, we focus our attention on a Fabry-P\'{e}rot configuration [see Fig. (\ref{ab}-b)] where the mode $\hat{a}_1$ is injected into a cavity. Such injection of a propagating optical mode into a cavity is standard in LIGO \cite{LIGO}, and in cavity-based quantum networks \cite{zoller1, Parkins, aspelmeyer}. Obviously, the injection itself entails a decoherence of the field, which we will take into account. 

After mode $\hat{a}_1$ is injected into the cavity it starts interacting with the mechanical oscillator. In a frame rotating at the frequency $\omega_1$ of mode $\hat{a}_1$, this interaction is  

\begin{equation}
\hat{H}_\mathrm{int} = \hat{b}^\dag\hat{b} - g \hat{a}^\dag_{1}\hat{a}_{1} (\hat{b}^\dag + \hat{b}),
\end{equation}

where $g = g_0/\omega_m$ is the scaled coupling parameter, $\omega_m$ is the angular frequency of the mechanical oscillator ($\hat{b}$), $g_0 = x_{\mathrm{zpf}} \omega_{1}/L $ is the radiation-pressure interaction strength, $L$ is the cavity length at equilibrium,  and $x_{\mathrm{zpf}}$ is the zero-point fluctuation amplitude (we set $\hbar = 1$) \cite{Bose1999}. Given the recent possibilities of ground state cooling \cite{zoller1,Romero-Isart,kiesel,Millen}  we will assume that the oscillator is initially in a coherent state $|\alpha\rangle$. The evolution in the absence of any source of decoherence can be solved straightforwardly \cite{bose, mancini}. In this ideal case, the dynamics is characterized by a displacement of the mirror position, conditioned on the photon number $n$:

\begin{eqnarray}
\nonumber \lambda^n\ket{n}_{1}\ket{\alpha} &\rightarrow& \lambda^n e^{ig^2n^2(t-\sin t)}e^{ign\mathrm{Im}[\alpha \eta]}\\
&\times& \ket{n}_{1}\ket{\alpha e^{-it} + gn\eta}.
\end{eqnarray}

Here $\ket{n}$ is a photonic Fock state, $\eta = 1 - e^{-it}$, and $t$ represents a scaled time, being the actual time multiplied by $\omega_m$.

However, in realistic conditions the state will be affected by decoherence. In order to give a full analytic solution, we assume that the cavity decay $\kappa$ is much smaller than the mechanical frequency $\omega_m$ (the resolved side-band regime already attained in several setups \cite{Kip, kiesel, Millen}). We solve the Markovian master equation at zero temperature for the decoherence of the oscillator following the procedure in the Appendix of Ref. \cite{bose}. In this case, the master equation reads as: 

\begin{equation}
\frac{d\hat{\rho}(t)}{dt}=-i[\hat{H}_\mathrm{int},\hat{\rho}(t)]+\frac{\gamma}{2}\left[2\hat{b}\hat{\rho}(t)\hat{b}^\dag - \hat{b}^\dag\hat{b}\hat{\rho}(t) - \hat{\rho}(t)\hat{b}^\dag\hat{b}\right],
\end{equation}

being $\gamma$ the mechanical energy damping rate. Another inevitable source of decoherence is the attenuation due to the injection of the light beam into the cavity. To model this, we consider a beam splitter (BS) in front of the fixed cavity-mirror, such that one port of the latter is fed with mode $\hat{a}_1$ and the other with a vacuum field \cite{kim}. Under these sources of decoherence, the analytic solution is

\begin{eqnarray}
\nonumber \hat{\rho}(t) &=& |1 - \lambda^2| \sum_{n,m=0}^\infty\sum_{k=0}^{\mathrm{min}[n,m]} G_{nm}^k(\theta) \mathcal{C}_{nm}^k e^{-D_{nm}^{\gamma,k}(t)}\\
&\times& \ket{n-k,n}\bra{m-k,m} \otimes \ket{\phi_n^k(\gamma,t)}\bra{\phi_m^k(\gamma,t)} \label{rho}
\end{eqnarray}

where the $\theta$ angle is related with the reflection coefficient of the BS as $r = \cos(\theta/2)$. The other terms are

\begin{eqnarray}
 \nonumber \mathcal{C}_{nm}^k &=& \lambda^{n+m}e^{ig^2[t - \sin t][(n-k)^2 - (m-k)^2]} e^{ig\mathrm{Im}[\alpha\eta](n - m)},\\
 \nonumber G_{nm}^k(\theta) &=& \sqrt{\binom{n}{k}\binom{m}{k}}\cos^{2k}\frac{\theta}{2}\sin^{n-k}\frac{\theta}{2}\sin^{m-k}\frac{\theta}{2},\\
 \nonumber \phi_n^k(\gamma, t) &=& \frac{ig(n-k)\left(1-e^{-(i + \gamma/2)t}\right)}{i + \gamma/2} + \alpha e^{-(i+\gamma/2)t},\\
 \nonumber D_{nm}^{\gamma,k}(t) &=& -\frac{\gamma}{2} \int_0^t \Big(|\phi_n^k(\gamma,t')|^2 + |\phi_m^k(\gamma,t')|^2\\
 &-& 2\phi_n^{*k}(\gamma,t')\phi_m^k(\gamma,t')\Big) dt'.
\end{eqnarray}

In the above, in order to keep analytic tractability, we have assumed that the light-mechanics coupling is absent during injection. This is actually feasible for a new class of optomechanical systems where a levitated trapped object inside a cavity embodies the mechanical element \cite{zoller1, Romero-Isart, Romero-Isart1}. Shifting the trapped position of this object with respect to the cavity field can change the strength of coupling \cite{kiesel-pvt-com, barker-pvt-com}. We still believe that our results reflect the general case (even when the light mechanics coupling is present during injection) well as we have modeled both phenomena separately. Moreover, a Q-switching of the cavity could be possible via a suitable intra-cavity scatterer as shown in Ref. \cite{2014Tommasso-ale}. Thus one can temporarily lower the \textit{Q} factor of the cavity during the injection --- in order to  get the light in much faster than the mechanical time scale --- and then rump it up again during the optomechanical evolution.  

\section{Entanglement Concentration}\label{sec:ent-conc}

In order to concentrate the entanglement in the initial TMSV, we proceed via measuring the quadrature position of the oscillator \cite{vanner} through an inefficient detector (modeled as an ideal detector preceded by a beam splitter of transmissivity $\nu$). This corresponds to the positive-operator valued measure (POVM) 

\begin{equation}
\hat{\Pi}(q) = \frac{1}{\sqrt{2\pi\delta_q^2}}\int_{-\infty}^\infty e^{-\frac{(q - y)^2}{2\delta_q^2}}\ket{y}\bra{y} dy,
\end{equation}

where $q=x \sqrt{m\omega_m/\hbar} $ is the dimensionless position of the oscillator (with actual position $x$), $m$ is the oscillator mass, and $\delta_q^2 = \frac{1-\nu}{4\nu}$ \cite{Alessandro-book}. The state (unnormalized) after the measurement, conditioned to an outcome $q$, is given by

\begin{eqnarray}
\nonumber \hat{\rho}(q)_{12} &=& \frac{|1 - \lambda^2|}{\sqrt{2\pi\delta_q^2}} \sum_{n,m=0}^\infty\sum_{k=0}^{\mathrm{min}[n,m]} G_{nm}^k(\theta) \mathcal{C}_{nm}^k e^{-D_{nm}^{\gamma,k}(t)}\mathcal{I}_{nm}^k\\
&\times& \ket{n-k,n}\bra{m-k,m} \label{rhos1s2}
\end{eqnarray}

where

\begin{equation}
\mathcal{I}_{nm}^k = \int_{-\infty}^\infty \psi_{\phi_n^k(\gamma,t)}(x) \psi_{\phi_m^k(\gamma,t)}^*(x) e^{-\frac{(q - x)^2}{2\delta_q^2}} dx
\end{equation}

in which $\psi_\xi(q) \equiv \langle q \ket{\xi}$ is the position wave-function of an arbitrary coherent state $\ket{\xi}$. The probability density function (PDF) of the outcome $q$ is

\begin{equation}
 p(q) = \frac{|1-\lambda^2|}{\sqrt{2\pi\delta^2_q}} \sum_{n=0}^\infty \sum_{k=0}^n \lambda^{2 n} G_{nn}^k(\theta) \mathcal{I}_{nn}^k
 \label{prob}
\end{equation}

To quantify the entanglement we use the negativity \cite{neg1, neg2}, defined as $N(t) = 1/2 \sum_i(|\varepsilon_i| - \varepsilon_i)$, where $\varepsilon_i$ are the eigenvalues of the partial transposition of the normalized version of $\hat{\rho}(q)_{12}$ of Eq. (\ref{rhos1s2}).

A numerical inspection of Eq. (\ref{prob}) reveals that a change in the initial amplitude from $|\alpha|e^{i\phi_\alpha}$ to $|\alpha '|e^{i\phi_\alpha '}$ entails a rigid shift of the outcome probability $p(q)$. Particularly, in absence of any source of decoherence, this rigid shift reads as $\Delta q = \sqrt{2}[|\alpha|\cos(\phi_\alpha - t) - |\alpha '|\cos(\phi_\alpha ' - t)]$. We verified numerically that also the entanglement negativity is subjected to the same shift, which implies that a change in $\alpha$ can be accounted for by selecting the measurement outcome $q$ accordingly. Given this, we set for the rest of this work the initial coherent state to $\alpha = 0$.

\begin{figure}
 \centering \includegraphics[width=\linewidth]{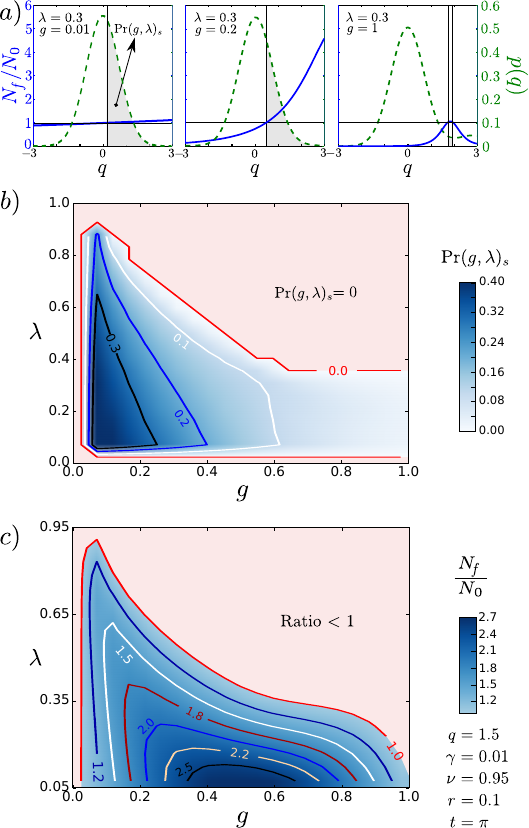} 
  \caption{Upper panel a): We plot in the left $y$-axis the ratio of the negativity $N_f/N_0$ as a function of the oscillator's position $q$ (solid line), where $N_f$ ($N_0$) stands for the distilled (initial) negativity. In the right $y$-axis we show the PDF as a function of $q$ (dashed line). In the middle panel b) we illustrate the concentration success probability ($\mathrm{Pr}(g,\lambda)_s$) corresponding to the shaded region in the upper panel. Finally, in the bottom panel c) we show the ratio of negativity as function of $\lambda$ and $g$ for a specific oscillator's position $q = 1.5$.}
\label{proglambda}
\end{figure}

We now have the ingredients to assess entanglement concentration. For a fixed set of values ($\gamma = 0.01, \delta_q \approx 0.11, r = 0.1, t=\pi$, $\nu = 0.95 \rightarrow \lambda = 0.3$) we plot in the left $y$-axis of Fig. (\ref{proglambda}-a) the ratio of the final ($N_f$) to initial ($N_0$) negativity (solid line) as a function of the outcome $q$ of the measurement of the oscillator position, where $N_0$ was computed considering the pure two-mode squeezed state before incoupling $\ket{\psi(0)}_{\mathrm{TMSV}} = \sqrt{1 - \lambda^2}\sum_{n=0}^{\infty} \lambda^n \ket{n,n}_{12}$. In the right $y$-axis, we show its corresponding PDF (dashed line) as a function of $q$. The success probability of the concentration protocol, namely the probability of obtaining $N_f > N_0$, is given by the shaded region and is:

\begin{equation}
 \mathrm{Pr}(g,\lambda)_s = \int_{N_f>N_0}p(q) dq.
\end{equation}

In Fig. (\ref{proglambda}-a) we illustrate three representative cases. For weak optomechanical coupling ($g=0.01$), one achieves a large success probability though at the cost of an almost negligible  increase in negativity $N_f \approx N_0$. For intermediate coupling ($g=0.2$) the negativity is significantly enhanced, still retaining a high success probability. On the other hand, for large coupling ($g = 1$), not only $N_f \lesssim N_0$ but also the success probability is considerably small. Thus an optimal region of the coupling value emerges, given that the entanglement concentration is predominantly achieved for \textit{intermediate} radiation-pressure coupling. Similarly we can also see that entanglement concentration is achieved for intermediate values of the initial entanglement, implying an optimal parameter region $0.2 \lesssim \{g,\lambda\} \lesssim 0.4$, for which $\mathrm{Pr}(g,\lambda)_s \gtrsim 0.2$ (see Figs. (\ref{proglambda}-b) and (\ref{proglambda}-c)).

The reason for this behaviour can be intuitively understood considering the structure of the TMSV state and its evolution under the concentration protocol. The states of the whole system (in absence of decoherence) before and after the optomechanical interaction are given by $\ket{0} \sum_n \lambda^n \ket{n,n}_{1,2}$ and $\sum_n \lambda^n e^{ig^2n^2\pi}\ket{2 g n }\ket{n,n}_{1,2}$ respectively. The states $\{\ket{2 g n }\}_n$ become more and more distinguishable for larger $g$. As a consequence, the measurement of the oscillator position effectively becomes a sharp measurement of Fock state inside the cavity \cite{fock-meter} that projects the two light beams into a factorized state $|n,n\rangle_{1,2}$. This intuitively explains the failure of the concentration protocol for large $g$. The failure for large $\lambda$ is instead due to the fact that the number of photon Fock states compatible with a specific outcome $q$ is finite (for any non-zero $g$). For large enough $\lambda$, this finite superposition of a small set of Fock states $|n,n\rangle_{1,2}$ is not enough to exceed the entanglement of the initial TMSV. Note that our results indicate that the concentration protocol is robust against large injection losses (in Fig. \ref{proglambda} we considered a beam-splitter reflectivity of $r=0.1$), which in turn suggests robustness against cavity and extraction losses as well.

\section{Quantum Teleportation with the Concentrated State}\label{sec-tel}

\begin{figure}
 \centering \includegraphics[width=\linewidth]{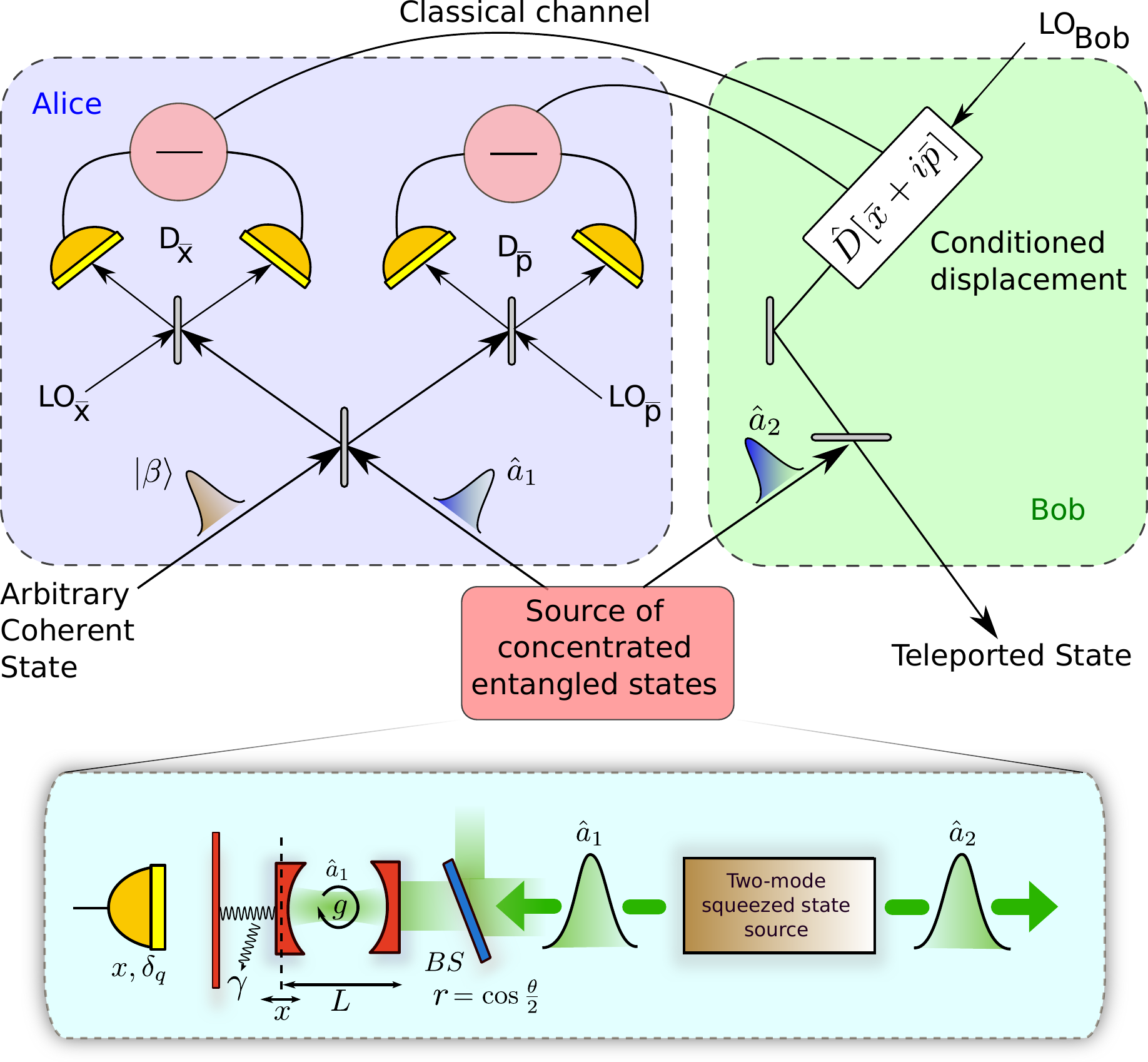} 
  \caption{Standard quantum teleportation procedure \cite{BK}. We later show in Fig. \ref{teleportation2} that, the teleportation of an arbitrary coherent state $\ket{\beta}$ is enhanced by using our concentrated state.}
\label{teleportation}
\end{figure}

For communication purposes, the ultimate application of entanglement concentration (and for that matter of entanglement distillation) is the enhancement of quantum teleportation fidelity. In the following we will show how the teleportation of an arbitrary coherent state $\ket{\beta}$ is enhanced by using our concentrated state. The full teleportation procedure is illustrated in Fig. \ref{teleportation}. First, state $\ket{\beta}$ is combined with mode $\hat{a}_1$ of the concentrated state $\hat{\rho}(q)_{12}$ into a balanced (50:50) beam splitter (BS). Subsequently, a measure of the position and momentum quadratures at the output ports of the BS are performed, resulting into outcomes $\bar{x}$ and $\bar{p}$ respectively. Lastly, to fully achieve the teleportation protocol, a displacement operation $D[\bar{x} + i\bar{p}]$ of the outcome state is realized.

Let us begin by combining the first light mode $\hat{a}_1$ of the concentrated state with the state to teleport ($\ket{\beta}$) into a 50:50 BS. To calculate this, we will refer to Ref. \cite{Agarwalbook}, where we can obtain straightforwardly the outcome of two modes states passing through a balanced BS in the Fock number basis. There, it is found that

\begin{equation}
 \ket{n,n'}_{out} = \sum_{j=0}^n\sum_{j'=0}^{n'} \mathcal{D}_{j,j'}(n,n')\ket{j+j',n+n'-j-j'}\label{bs}
\end{equation}

where for a 50:50 BS,

\begin{eqnarray}
 \nonumber \mathcal{D}_{j,j'}(n,n') &=& \binom{n}{j}\binom{n'}{j'}\left(\frac{1}{\sqrt{2}}\right)^{n'+n}(-1)^{n-j}\\
 &\times&\sqrt{\frac{(j+j')!(n+n'-j-j')!}{n!n'!}}.
\end{eqnarray}

Thus, after writing down the concentrated state in Eq. (\ref{rhos1s2}) in conjunction with the coherent state to teleport ($\mathrm{tel}$), we can easily calculate the state after passing through the BS using Eq. (\ref{bs}) as following

\begin{eqnarray}
\nonumber \hat{\rho}_{1,tel,2} &=& \frac{|1 - \lambda^2|e^{-|\beta|^2}}{p(q)\sqrt{2\pi\delta^2_\nu}}\sum_{n,m=0}^\infty\sum_{k=0}^{\mathrm{min}[n,m]}G_{nm}^k(\theta)\mathcal{C}^k_{nm}\\
 \nonumber &\times& e^{-D_{nm}^{\gamma,k}(t)} \mathcal{I}_{nm}^k\sum_{n',m'=0}^\infty\frac{\beta^{n'}\beta^{*m'}}{\sqrt{n'!m'!}}\\
 \nonumber &\times& \sum_{j=0}^{n-k}\sum_{j'=0}^{n'} \mathcal{D}_{j,j'}(n-k,n') \sum_{l=0}^{m-k}\sum_{l'=0}^{m'} \mathcal{D}_{l,l'}(m-k,m')\\
 \nonumber &\times& \ket{j+j',n-k+n'-j-j'}_{1,\mathrm{tel}}\\
 &&\bra{l+l',m-k+m'-l-l'}_{1,\mathrm{tel}}\otimes\ket{n}_{2}\bra{m}_{2}. \label{1anc2}
\end{eqnarray}

Now, we proceed to measure the position (momentum) quadrature of the transmitted (reflected) beam in Eq. (\ref{1anc2}), obtaining the following unnormalized state

\begin{widetext}
\begin{eqnarray}
 \nonumber &&\hat{\rho}_{2} = \frac{|1 - \lambda^2|e^{-|\beta|^2}}{p(q)\sqrt{2\pi\delta^2_\nu}}\sum_{n,m=0}^\infty\sum_{k=0}^{\mathrm{min}[n,m]}G_{nm}^k(\theta) \mathcal{C}^k_{nm} e^{-D_{nm}^{\gamma,k}(t)} \mathcal{I}_{nm}^k \sum_{n',m'=0}^\infty\frac{\beta^{n'}\beta^{*m'}}{\sqrt{n'!m'!}} \sum_{j,j'=0}^{n-k,n'}\mathcal{D}_{j,j'}(n-k,n')\\
 &\times&\bra{\bar{p}} j + j' \rangle  \bra{\bar{x}} n - k + n' - j - j' \rangle \sum_{l,l'=0}^{m-k,m'}\mathcal{D}_{l,l'}(m-k,m') \bra{\bar{p}}l + l'\rangle ^*   \bra{\bar{x}} m-k + m' - l - l' \rangle \ket{n}_{2}\bra{m}_{2},\label{notnormteleport}
\end{eqnarray}
\end{widetext}

where 

\begin{eqnarray}
 \bra{\bar{x}}n\rangle &=& \left(\frac{1}{\pi}\right)^\frac{1}{4}\frac{1}{\sqrt{2^nn!}}e^{-\bar{x}^2/2}H_n(\bar{x}),\\
 \bra{\bar{p}}n\rangle &=& (-i)^n\left(\frac{1}{\pi}\right)^\frac{1}{4}\frac{1}{\sqrt{2^nn!}}e^{-\bar{p}^2/2}H_n(\bar{p})
\end{eqnarray}

and $H_n(x)$ is the Hermite polynomial of degree $n$.

The success of the teleportation protocol for an arbitrary coherent state $|\beta\rangle$ and a set of measurement outcomes $\{\bar{x}, \bar{p}\}$ can be quantified by the Fidelity function: 

\begin{equation}
f_\beta(\bar{x},\bar{p}) = \bra{\beta}\hat{D}[z]\frac{\hat{\rho}_2}{p_\beta(\bar{x},\bar{p})}\hat{D}^\dagger[z]\ket{\beta} \;,
\end{equation}

where $p_\beta(\bar{x},\bar{p})$ denotes the outcome probability and the actual values of the measurement outcomes are used to correct Bob's state via the displacement $\hat{D}[z] = D[\bar{x} + i\bar{p}]$. Averaging over all possible outcomes the fidelity function reads as:

\begin{eqnarray}
 \langle \mathcal{F}\rangle_\beta &=& \int_{-\infty}^{+\infty}f_\beta(\bar{x},\bar{p}) p_\beta(\bar{x},\bar{p})  d\bar{x}d\bar{p},\\
 &=& \int_{-\infty}^{+\infty} \bra{\beta - \bar{x} - i\bar{p}}\hat{\rho}_2 \ket{\beta- \bar{x} - i\bar{p} }   d\bar{x}d\bar{p} \label{qtele}\;.
\end{eqnarray}

Similarly to what happens for the case of teleportation using a TMSV, numerical evaluations of Eq. (\ref{qtele}) show no dependence on the coherent state to teleport, i.e. $\langle \mathcal{F}\rangle_\beta \rightarrow \langle \mathcal{F}\rangle$. 

To contrast the performance of the quantum teleportation protocol described above, we proceed to compute the ratio of the Fidelity $\langle \mathcal{F}\rangle$ against the well-known result \cite{Alessandro-book} of the Fidelity for an initial TMSV state, as shown in Eq. (\ref{tmsv})

\begin{equation}
\langle \mathcal{F}_{\mathrm{TMSV}} \rangle = \frac{1 + \lambda}{2}.
\end{equation}

\begin{figure}
 \centering \includegraphics[width=\linewidth]{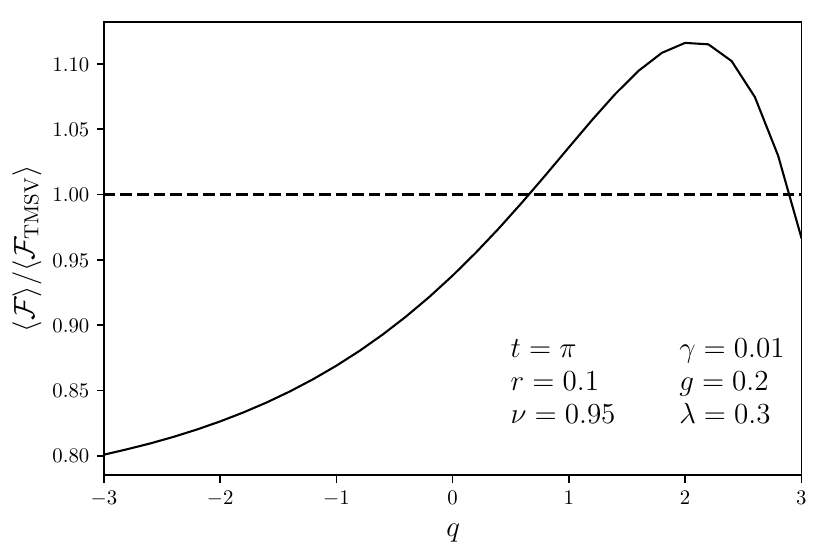} 
  \caption{Ratio between the Fidelity $\langle \mathcal{F} \rangle$ of the teleportation protocol that uses as a resource the concentrated state $\hat{\rho}(q)_{12}$ [in Eq.(\ref{rhos1s2})] and the Fidelity  $\langle \mathcal{F}_{\mathrm{TMSV}} \rangle$ obtained when an ideal TMSV is used. The ratio is plotted as a function of the outcome $q$ of the oscillator's position measurement. Other parameters are $\gamma = 0.01, \nu = 0.95, r = 0.1, t = \pi$, $\lambda = 0.3$, and $g = 0.2$.}\label{teleportation2}
\end{figure}

In Fig. \ref{teleportation2} we illustrate $\langle \mathcal{F} \rangle/\langle \mathcal{F}_{\mathrm{TMSV}} \rangle$ as a function of the oscillator's position $q$. We have considered the optimal parameters as in Fig. (\ref{proglambda}-b), i.e. $\gamma = 0.01, \delta_q \approx 0.11, r = 0.1, t = \pi$, $\nu = 0.95 (\lambda = 0.3)$, and $g = 0.2$. As seen, for the same values of the oscillator's position for which we can concentrate the initial TMSV state $q \gtrsim 0.5$, we are also able to enhance the quantum teleportation Fidelity on average.

\section{Enabling of Entanglement Distillation}\label{sec-enabling}

Having shown how our entanglement concentration procedure can be used to enhance a pivotal quantum communications protocol, we now outline how scalable entanglement distillation (SED) procedures can also be constructed. Note that there exist SED schemes that only rely on the repeated production (supply) of a fixed non-Gaussian entangled state $|\psi^0\rangle$ and linear optics \cite{Datta}. Let us briefly examine the principal steps of the procedure described in Ref. \cite{Datta}. Essentially, the authors propose a so-called ``entanglement distillery'', where a vacuum-conditioned iterative process leads to a highly distilled state. In particular, the scheme is based solely on two main steps and feasibly implemented using four quantum memories (a novel fixed hardware that plays the role both for storage, as well as processing the quantum information). Firstly, the ``malting'' step generates a weak entangled state ($\ket{\psi^0}$), one to be used as a quantum state provider in a later stage, to be more precise, an optical non-Gaussian resource denoted as $\ket{\psi^0} = \sum_n \alpha_n^0 \ket{n}\ket{n}$ in Schmidt decomposition. Secondly, two copies of $\ket{\psi^0}$ are distributed between two parties and combined into a 50:50 BS, subsequently, a vacuum detection on one of the emerging modes of the BS heralds the success of the protocol and thus giving rise to a more entangled state, let say $\ket{\psi^1}$. Following the above ``mashing'' steps (BS mixing and vacuum detection), the resulting state $\ket{\psi^1}$ (from the previous iteration) in conjunction with a fresh copy of $\ket{\psi^0}$ are now combined into a BS and vacuum conditioned as before to produce a more entangled state $\ket{\psi^2}$, etc. As stated in Ref. \cite{Datta}, the final amount of entanglement is governed by the initial states (resource supplied in the ``malting'' steps) and the number of iterations (``mashing'' steps). Our procedure provides precisely such a resource at a rate $\sim \omega_m$, as for optimal $g$, for any measurement outcome $q$, a non-Gaussian entangled state is produced. In essence the probabilistic step of photon-subtraction in the usual schemes (the ``malting'' step in Ref.\cite{Datta}) can be replaced by our protocol as the provider of $\ket{\psi^0}$, while the rest of the entanglement distillation protocol (``mashing'' steps) remain unchanged. As opposed to photon subtraction, here we get a non-Gaussian state with ``unit probability''. Although the state will be a different state $|\psi_0(q)\rangle$ for each outcome $q$, a curious fact is that over any small window $\Delta q$ of possible $q$, the resulting states are nearly the \textit{same} because of the coarse grained nature of the effective measurement in the cavity Fock basis. For example for $\Delta q \sim 0.5$,  $|\psi_0(q)\rangle$ remains the same state with $\sim 0.98$ average fidelity  --- see plots in Appendix A. Thereby, the whole domain of possible $q$, say $-3$ to $+3$ could be divided to bins of width $\Delta q \sim 0.5$ and, for example, $12$ entanglement distillation recursions can be carried out in parallel for outcomes in each bin. In this example, one would need to have 12 cavity-TMSV systems operated in parallel so that the $|\psi_0(q)\rangle$ emerging from each of these systems can, on average, feed in as the $\ket{\psi^0}$ for one of the parallel recursive protocols of distillation.

\section{Experimental feasibility}\label{sec-exp}

The generation of the TMSV state with $0 < \lambda < 0.5$ is routine. Additionally, our protocol requires the initialization of the mechanical oscillator in an arbitrary pure coherent state. For on-chip integrated opto-mechanical systems, this has already been achieved at room temperatures using laser cooling \cite{chan, Aspel-new}, in much of which the resolved side-band regime is also already achieved. Moreover, our optimal $g \sim 0.2$ (moderately strong) for the optomechanical interaction has already been achieved/exceeded/quoted in the same \cite{Krause} and similar systems \cite{murch,vanner}. Modest improvements in some on-chip systems \cite{Painter, chan} (e.g., by decreasing both $\omega_m$, as $g \propto \omega_m^{-\frac{3}{2}}$, and $\kappa$ by just 1-2 orders of magnitude) are required, while exciting recent proposals to exceed the required $g$ have been made \cite{xuereb, Barclay, Romero-Isart}. The key stage of this work consists in measuring the oscillator. After the pulse $\hat{a}_1$ interacts with the oscillator, a second auxiliary pulse with a duration much smaller than $1/\omega_m$ is injected into the cavity. The optical phase of the emerging field (correlated with the mechanical position) is then measured via balanced homodyne detection \cite{vanner}. 

Another feasibility front opened up in very recent experiments are trapped/levitated systems which are tantalizingly close to be cooled to the ground state by feedback/laser cooling at room temperature with necessary rates already achieved \cite{kiesel, Millen, Hendrik}. In the same systems, position measurements of the mechanical oscillator have been demonstrated to be highly precise $10^{-15}/\sqrt{\mathrm{Hz}}$ \cite{Raizen,Novotny,Hendrik} without hurdles. Moreover, ideally, we want our optomechanical coupling to cease  after the measurement as we do not want entanglement of the distilled quantum state with mechanics while it is being used for some other protocol. This can be achieved, for example, by using trapped mechanical objects as oscillators interacting with a cavity field (for which $g\sim 0.1$ should be possible from the parameters of Ref.\cite{zoller1} by decreasing the cavity waist to $5 \mu m$): an optically trapped object can be pushed away by a strong pulse \cite{kiesel-pvt-com}; a charged object held in a Paul trap can be suddenly shifted relative to the cavity field by suddenly shifting the trap centre \cite{barker-pvt-com}. 

\section{Concluding Remarks}\label{sec-conclusions}

We have presented a first application of the bare (non-linear) radiation-pressure coupling in a practical quantum communication protocol, showing how optomechanics can fill a crucial gap in enabling long distance quantum networks using Gaussian states of light. Our proposal uses an indirect measurement of the photon number of the field inside a cavity through the position measurement of a mechanical element coupled to it.  For an optimal strength of the coupling, the photon number is measured weakly or ``unsharply'' and this results in entanglement concentration or producing a non-Gaussian resource state $\ket{\psi^0}$ for scalable entanglement distillation. As opposed to photon-photon nonlinearities, the radiation pressure coupling is a direct matter-light interaction and hence less noisy. Moreover optomechanical systems, especially the zipper cavity arrays in crystals are highly integrable to allow many processes in parallel as is necessary for entanglement distillation. For a vacuum state inside the cavity, the position meter does not move, corresponding to a failed outcome of distillation. Thus our procedure has a degree of similarity with the known purely optical procedure of photon subtraction \cite{Cirac, Fiurasek, browne, Walmsley,Opatrny,Olivares}, where also the vacuum component is filtered, and we have comparable figures of merit. 

The state obtained through our protocol is always non-Gaussian, and thus it can serve as the malting step of scalable entanglement distillation \cite{Datta} ---or, more in general, for quantum computation purposes \cite{Menicucci,Datta2}. Moreover, the procedure here outlined could be useful also in a quantum repeater scenario for long distance communication, considering that there no further extraction of the distilled state from the optical cavity is needed.

\section*{ACKNOWLEDGMENTS}

V.M. acknowledges support from CONICYT Becas Chile 72110207 and from the Chinese Postdoctoral Science Fund 2018M643435. A.F. acknowledges funding from the John Templeton Foundation (Grant No. 43467). S.B. acknowledges the EPSRC Grant No. EP/J014664/1. We have benefited from comments of M. Vanner, G. J. Milburn, N. Kiesel, A. Rahman, P. F. Barker, D. E. Browne and particularly A. Datta.

\appendix

\section{Enabling of Entanglement Distillation}

\begin{figure}[b]
 \centering \includegraphics[width=\linewidth]{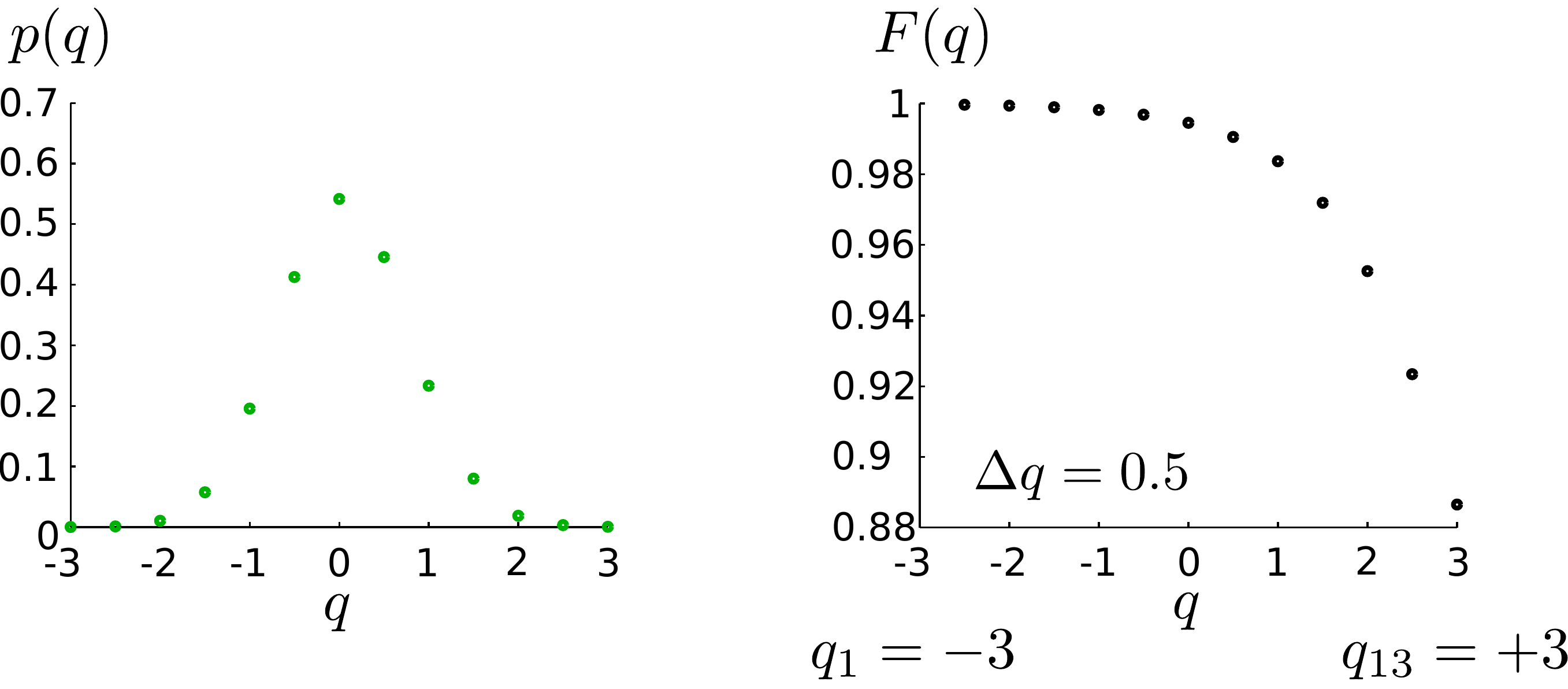} 
  \caption{Right panel shows the fidelity between two \textit{adjacent} concentrated states after measuring the oscillator's position, we have fixed $\Delta q = 0.5$. Left panel illustrates the corresponding PDF for the position measurement. Other parameters are $\gamma = 0.01, \nu = 0.95, r = 0.1, t = \pi$, $\lambda = 0.3$, and $g = 0.2$.} \label{fig2}
\end{figure}

From the main text, we have discussed that a non-Gaussian entangled state is generated at every $q-$measurement step ---as opposed from the probabilistic nature in photon-substraction schemes. Furthermore, although the state will be a different one $|\psi_0(q)\rangle$ for each outcome $q$ (being $|\psi_0(q)\rangle$ the provider state in the ``malting'' step \cite{Datta}), for a small window $\Delta q$ of possible $q$, the resulting states are nearly the \textit{same}.

To prove the above statement, we compute the average of fidelity

\begin{equation}
 \langle F \rangle = \frac{\sum_{i=1}^{12} \mathrm{p}(q_{i+1})F(q_{i+1})}{\sum_{i=1}^{12} \mathrm{p}(q_{i+1})}.
\end{equation}

$F(q_{j+1})$ correponds to the fidelity between two concentrated states, defined simply as

\begin{equation}
 F(q_{i+1}) = \mathrm{Tr}[\hat{\rho}(q_i)_{12}\hat{\rho}(q_{i+1})_{12}].
\end{equation}

where $\hat{\rho}(q)_{12}$ is given in Eq. (\ref{rhos1s2}), and
 $q_{j+1}$ and $q_j$ are two adjacent position measurements for the mechanical object, such as $\Delta q = q_{i+1} - q_i$. 

In the right panel of Fig. \ref{fig2}, we illustrate that $|\psi_0(q)\rangle$ in fact remains the same state with a fidelity above $0.89$, in the whole domain of possible $q = \{-3, +3\}$ ---where, in the left panel of Fig. \ref{fig2} we show its corresponding probability. For a window of $\Delta q = 0.5$, the entanglement distillation recursions can be carried out in parallel for outcomes in each bin. From the above values, it is straightforward to compute the average of the fidelity, which for this case reads as $\langle F \rangle \approx 0.98$.

\end{document}